\documentclass[journal]{IEEEtran}
\usepackage{amsmath,amsthm}
\usepackage{amsfonts}
\usepackage{amssymb}
\usepackage{multirow}
\usepackage[pdftex]{graphicx}
\usepackage{cite}
\usepackage{hyperref}
\hypersetup{colorlinks=true,urlcolor=red}

\newcommand{\Xc}{\mathcal{X}}
\newcommand{\Yc}{\mathcal{Y}}

\begin{document}
\title{Unsupervised Denoising for Satellite Imagery  using Wavelet Subband CycleGAN}

\author{Joonyoung~Song, Jae-Heon~Jeong, Dae-Soon~Park, Hyun-Ho~Kim, Doo-Chun~Seo,
 Jong~Chul~Ye,~\IEEEmembership{Fellow,~IEEE}% <-this % stops a space
\thanks{
This research was supported by ‘Satellite Information Application’ of the Korea Aerospace Research Institute (KARI).}% <-this % stops a space
\thanks{J. Song and J. C. Ye are with the Department of Bio and Brain Engineering,
Korea Advanced Institute of Science and Technology (KAIST), Daejeon 34141,
South Korea (e-mail: songjy18@kaist.ac.kr; jong.ye@kaist.ac.kr).}
\thanks{J. Jeong, D. Park, H. Kim, and D. Seo are with Image Data System Development Division,
Satellite Information Center, Korea Aerospace Research Institute (KARI), Daejeon 34133,
South Korea (e-mail: jjh583@kari.re.kr; parkds@kari.re.kr; kimhh@kari.re.kr; dcivil@kari.re.kr).}}

\markboth{Journal of \LaTeX\ Class Files,~Vol.~x, No.~x, xxx~2020}%
{SONG \textit{et al.}: Unsupervised Denoising for Satellite Imagery using Wavelet Subband CycleGAN}

\maketitle
\begin{abstract}
Multi-spectral satellite imaging sensors acquire various spectral band images such as red (R), green (G), blue (B), near-infrared (N), etc. Thanks to the unique spectroscopic property of each spectral band with respective to the objects on the ground, multi-spectral  satellite imagery can be used for various geological survey applications. Unfortunately,  image artifacts from  imaging sensor noises often affect the quality of scenes and have  negative impacts on the applications of satellite imagery. Recently, deep learning approaches have been extensively explored for the removal of noises in satellite imagery. Most deep learning denoising methods, however, follow a supervised learning scheme, which requires matched noisy image and clean image pairs that are difficult to collect in real situations.
% Also, most deep learning based methods for denoising are designed in the image domain, which leads to degradation of edges and detail information.
 In this paper, we propose a novel unsupervised  multi-spectral denoising method for satellite imagery using wavelet subband cycle-consistent adversarial network (WavCycleGAN). The proposed method  is based on unsupervised learning scheme using adversarial loss and cycle-consistency loss to overcome the lack of paired data. Moreover,  in contrast to the standard image-domain cycleGAN, 
 we introduce a wavelet subband domain learning scheme for effective denoising without sacrificing high frequency components such as edges and detail information. Experimental results  for the removal of vertical stripe and wave noises in satellite imaging sensors 
 demonstrate that the proposed method effectively removes noises and preserves important high frequency features of satellite images.

\end{abstract}


% \fi
\IEEEpeerreviewmaketitle

\section{Introduction}
\IEEEPARstart{M}{ultispectral} imaging sensors from a satellite capture different types of spectral band informations.
For instance, a typical high-resolution satellite has several imaging sensors
for multi-spectral bands such as red (R), green (G), blue (B),  near-infrared (N), etc. % band images.
 Each spectral band signal has unique spectroscopic characteristics, resulting in a variety of remote sensing applications such as agricultural planning \cite{Reference:mulla2013twenty}, traffic monitoring \cite{Reference:larsen2009traffic}, city planning \cite{Reference:pham2011case}, disaster analysis \cite{Reference:voigt2007satellite}, etc.

\begin{figure}[!t]
\centering
\includegraphics[width=0.9\linewidth]{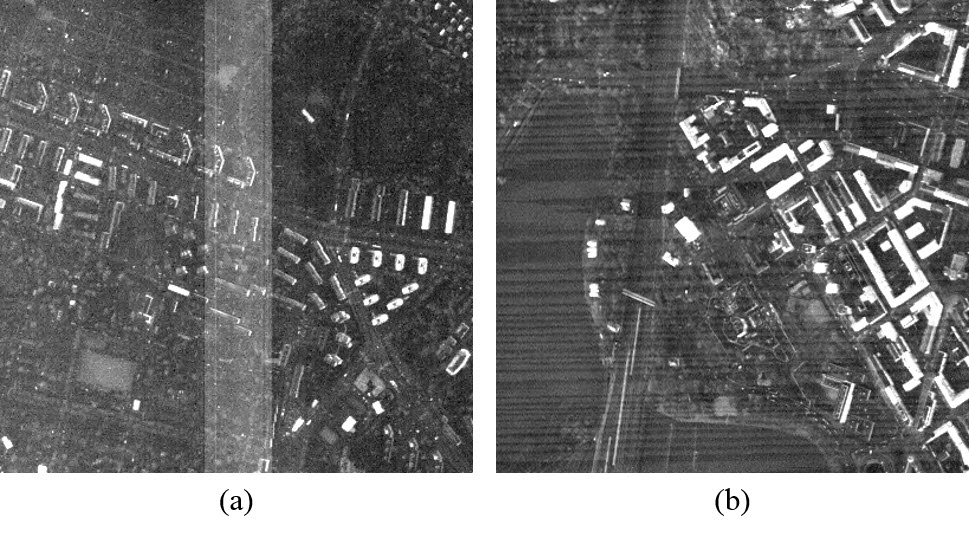}
\vspace{-0.3cm}
\caption{Examples of satellite images with (a) vertical stripe noise, and (b) wave noise.}
\label{figure1}
\end{figure}

Unfortunately, the quality of satellite images are often affected by various noise sources such as system calibration error, intrinsic properties of the hardware, sensitivity of the sensors, photon effect, and thermal noise. Fig.~\ref{figure1} shows typical examples of structured noise patterns in images from a high-resolution satellite such as vertical stripe noises and wave noises. The main cause of vertical stripe noises is an interference from the different scan timings of multi-spectral imaging sensors in a push broom scanner. Different sampling timings, and also the sensitivity of sensors, induce a different offset in each detector and generate vertical stripe noise patterns. Horizontal wave noise is an irregular wave noise pattern that is caused by interference from various hardware components.
Noises in images degrade the quality of the scenes and limit the use of satellite imagery. Therefore, one of the most important pre-processing for satellite images is the elimination of image noises that occur during the image acquisition process. 

Previously, various methods have been proposed for the removal of noise in satellite images. Conventional denoising methods follow model-based approaches using hand-crafted features and prior knowledge of data \cite{Reference:yuan2012hyperspectral, Reference:chang2015anisotropic, Reference:zhang2013hyperspectral, Reference:he2015hyperspectral, Reference:he2015total, Reference:du2016pltd, Reference:wu2017structure, Reference:wang2017hyperspectral, Reference:fan2018spatial}. 
However, the limitation of traditional model-based methods is a degradation in performance if predefined features or prior knowledges of the model do not fully reflect the properties of new data. 

Recently, deep convolutional neural network (CNN) have shown extraordinary performance in the image denoising problem \cite{burger2012image,mao2016image,Reference:zhang2017beyond,Reference:zhang2018ffdnet}. The advantage of using deep learning methods comes from the data-driven nature that automatically learns the optimal features for the task from the data.
In remote sensing applications, CNN-based denoising algorithms have been proposed and shown promising results \cite{Reference:yuan2018hyperspectral, Reference:chang2018hsi, Reference:zhang2019hybrid, Reference:guan2019wavelet, Reference:liu20193, Reference:shan2019hyperspectral, Reference:chang2019toward}. 
However, most CNN-based denoising methods for the satellite images are trained in a supervised manner. Supervised learning scheme requires structurally matched noisy image and clean target image pairs, which are difficult to obtain in real situations. 

To utilize unmatched image pairs, unsupervised learning methods should be used. 
Among the various approaches for unsupervised learning,
generative adversarial network (GAN) \cite{Reference:goodfellow2014generative} was proposed as  a distribution matching scheme so that it learns the distribution of the target domain from the input distributions.
However, the standard GAN approaches
often suffer from mode-collapsing behavior, which often generates artificial features.
To address the mode-collapsing problem,  unsupervised image-to-image translation using the cycle-consistent adversarial network (CycleGAN) was proposed \cite{Reference:zhu2017unpaired}. Specifically, the network is  trained in an unsupervised manner using  generative networks, and the cyclic consistency alleviates the generation of artificial features due to the mode collapsing problem of GAN. Inspired by the success of cycleGAN, Kang \textit{et al.} \cite{Reference:kang2019cycle} proposed cycle-consistent adversarial denoising network for multiphase coronary computed tomography (CT) angiography.
Here,  the denoising problem was considered as the image-to-image translation problem  between  two domains: noisy image domain and clean image domain, and the results by  Kang \textit{et al.} \cite{Reference:kang2019cycle} shows that
cycleGAN is a promising tool for unsupervised denoising.

Another limitation of most CNN-based denoising algorithms for satellite imagery is that the methods are designed in the image domain, %with mean squared error as a reconstruction loss, 
which  often leads to  blurred output and  loss of the edges and details information especially when the training data is not sufficiently many. High frequency components in satellite images contain important information that is crucial for the use of the satellite. Therefore, a desirable denoising algorithm should only remove noise components while preserving image details. Transform domain learning approaches are an alternative to image domain denoising methods. For instance, 
%wavelet transform can be utilized for the denoising problem. 
the advantage of using wavelet transform is that the image can be decomposed to directional subbands that can be used effectively to remove noise while preserving high frequency components. Based on this observation,  a denoising method based on the  wavelet domain deep learning was proposed in a supervised learning framework, which effectively removes noises without affecting image details \cite{Reference:guan2019wavelet, Reference:kang2017deep}.

Inspired by these approaches, here we propose a unsupervised multi-spectral denoising method for satellite imagery using wavelet subband cycle-consistent adversarial network (WavCycleGAN), and demonstrate its superior performance 
 for the removal of two structured noise patterns: vertical stripe noise and wave noise. 
Specifically, 
%we design an unsupervised denoising network with cyclic consistency to overcome the problem of unmatched image pairs for the satellite images. Furthermore, we propose a wavelet subband image learning scheme to remove structured noise patterns without sacrificing high frequency components of images. 
based on the property of target noises, specific wavelet subbands that contain majority of noises are selected for the wavelet recomposition to obtain the wavelet subband image. Then, %our cycleGAN is designed to the 
our cycleGAN network is trained in an unsupervised manner to learn the distribution matching between two wavelet subband domains from clean and noisy images, respectively.
% we can effectively reconstruct directional components of noise patterns
  
  Our experimental results show that our multi-spectral denoising method using WavCycleGAN effectively removes vertical stripe noise and wave noise  while preserving edges and details of images.

\section{Related Works}

\subsection{Model based Approaches}
%Many reconstruction algorithms have been proposed for denoising satellite images in the past years. Typically, 
Conventional satellite image denoising methods typically exploit model-based approaches which utilize hand-crafted representation and intrinsic properties of satellite images.
% Most traditional methods follow the optimization scheme, which solves certain problems with predefined constraints based on prior knowledge.

Satellite images tend to be piecewise smooth in the spatial domain \cite{Reference:he2015total}. Total variation (TV) denoising model \cite{Reference:rudin1992nonlinear} have been applied to the noise removal of satellite imagery because it effectively preserves high frequency information and enforces piecewise smoothness \cite{Reference:yuan2012hyperspectral, Reference:chang2015anisotropic, Reference:he2015total}. Yuan \textit{et al.} \cite{Reference:yuan2012hyperspectral} extended TV model to the spectral-spatial adaptive TV denoising model which considers spectral and spatial information of images.
Chang \textit{et al.} \cite{Reference:chang2015anisotropic} proposed the image destriping method using the anisotropic spectral-spatial total variation model. He \textit{et al.} \cite{Reference:he2015total} regularized their model with TV to enforce piecewise smoothness of images.
Clean satellite images that consist of multi-spectral images can be considered to have low-rank property \cite{Reference:zhang2013hyperspectral}. Based on the intrinsic sparsity of satellite images, low-rank matrix recovery (LRMR) approaches have been applied to noise removal problems in remote sensing \cite{Reference:zhang2013hyperspectral, Reference:he2015hyperspectral, Reference:he2015total}. 
Zhang \textit{et al.} \cite{Reference:zhang2013hyperspectral} introduced an image restoration method using LRMR. 
He \textit{et al.} \cite{Reference:he2015hyperspectral} proposed noise-adjusted framework that takes into account different properties of noises in different bands.
He \textit{et al.} \cite{Reference:he2015total} introduced a total variation-regularized low-rank matrix factorization (LRTV).
Recent works exploit tensor-based approaches with low-rank property and TV regularization to utilize spatial-spectral correlations of multi-spectral satellite imagery \cite{Reference:du2016pltd, Reference:wu2017structure, Reference:wang2017hyperspectral, Reference:fan2018spatial}.

However, the drawback of these model-based image restoration is the use of hand-crafted features and data model, which may degrade the performance of the algorithm if the image data has unexpected properties beyond their assumptions.
%The alternative to conventional model-based restoration can be a learning-based method which extracts features based on training data.

\subsection{Deep Learning  Approaches}
%Deep learning has recently been recognized as an effective tool for image restoration because it has led to significant performance improvements in many computer vision problems such as denoising \cite{Reference:zhang2017beyond}, \cite{Reference:zhang2018ffdnet}, super-resolution \cite{Reference:ledig2017photo}, and deblurring \cite{Reference:kupyn2018deblurgan}. Also, 
In the field of remote sensing, many deep learning based methods have been proposed for the removal of noise in satellite imagery. 
Yuan \textit{et al.} \cite{Reference:yuan2018hyperspectral} proposed a spatial-spectral deep residual learning method using CNN for hyperspectral images (HSID-CNN). Their method utilized spatial and spectral information by using noisy input and adjacent spectral bands. 
Chang \textit{et al.} \cite{Reference:chang2018hsi} proposed a method for hyperspectral image denoising via CNN (HSI-DeNet). HSI-DeNet consists of dilated convolution layers and exploits residual learning approach to effectively remove noise.
Zhang \textit{et al.} \cite{Reference:zhang2019hybrid} introduced a spatial-spectral gradient network (SSGN) for the removal of hybrid noise in hyperspectral images. SSGN use spatial and spectral gradient information to extract important features of satellite images. 
Guan \textit{et al.} \cite{Reference:guan2019wavelet} proposed wavelet deep neural network for stripe noise removal. They trained the network in wavelet domain for the effective denoising.
Other deep learning based denoising method for satellite imagery can be found in \cite{Reference:liu20193, Reference:shan2019hyperspectral}, and \cite{Reference:chang2019toward}.

Most deep learning based image restorations for satellite imagery follow a supervised learning scheme. A supervised learning method requires a paired dataset consisting of a noisy image and spatially matched clean image to train the network. The need for a paired dataset, however, limits the use of a supervised learning scheme in practice, since paired satellite images are difficult to collect in real situations. To mitigate this problem, many researchers added synthesized noise to relatively clean images. Although the trained model using synthetic noise works well for artificial noise, it is difficult to estimate real noise components that are complex in practice.

\subsection{Our Contributions}

\subsubsection{Unsupervised Learning Approach}

Although it is difficult to obtain matched clean and noisy image pairs from satellite imagery,  it is much easier to obtain {\em unmatched} clean and noisy image
data sets in practice. This is because in some situations sensors are affected less by the noises, or the assumption of the model-based approaches are sufficiently accurate to generate clean images. However, the practical issue is that these clean images are not matched to the noisy multi-spectral image data that one is interested in processing.

In this scenario, an unsupervised learning scheme that uses the unmatched clean and noisy image data set is a perfect fit.
Therefore, one of the most important contributions of our paper is an unsupervised learning scheme  that uses the unmatched
clean and noisy images for neural network training.
%
%using cycle-consistent generative adversarial network to address this problem.
%%Therefore, 
%%were trained in a supervised manner, 
%our method employs an unsupervised learning scheme that does not require paired datasets. Thus, in real situations, our unsupervised learning approach is more practical than the existing methods designed in supervised learning approach. 
Specifically, to train the network in an unsupervised way, we use adversarial loss and cycle-consistency loss. Accordingly, noise patterns can be removed efficiently without requiring the matched data set.

\subsubsection{Wavelet Subband Learning}
Typically, deep learning based image restorations in remote sensing are designed in the image domain. % with the mean square error as a loss function. 
Unfortunately, perfect noise separation using image domain deep network  is often difficult especially when enough training data sets are not available. Thus, edges and detail information are often removed by the denoising networks that are  applied directly in the image domain.
% In addition, using a mean square error as a restoration loss produces a smoothed output, and may also remove high frequency components that are important for the application of satellite imagery.

Unlike the existing deep learning based algorithms designed in the image domain, we train our model using wavelet subband images that are obtained from subset of wavelet bands containing noises.   Accordingly, the spectral contents in the other bands are not altered by the neural network so that we can achieve efficient noise removal without sacrificing high frequency information.
%the image blurring 
%
%Our experimental results show that the use of wavelet subband image improves the denoising performance.

\section{Theory}

\subsection{Wavelet Subband Image}
As described before, one of the disadvantages of the image domain deep learning is that  output images from neural networks tend to be blurry since  high frequency components such as edges and details of images can be altered by the reconstruction algorithm. To remove noise patterns while preserving image details, here we propose a wavelet subband  deep learning.

The procedure for generating wavelet subband images is as follows. First, we used the 2D Daubechies-3 wavelet transform (db3) to decompose the input image to subband images such as approximation (LL), horizontal detail (LH), vertical detail (HL), and diagonal detail (HH) bands. With $K$-th level wavelet decomposition, we have $\{LL_{i}\}_{i=1}^{K}$, $\{LH_{i}\}_{i=1}^{K}$, $\{HL_{i}\}_{i=1}^{K}$, and $\{HH_{i}\}_{i=1}^{K}$ subband images.
The advantage of using the wavelet transform is that we can decompose an image into directional subbands.
Therefore, if the noises have specific directional properties, % components,
the noises can be usually confined in specific subset of wavelet bands.
This is the prior information we want to explore in designing the neural network.

\begin{figure}[!t]
\centering
\includegraphics[width=0.9\linewidth]{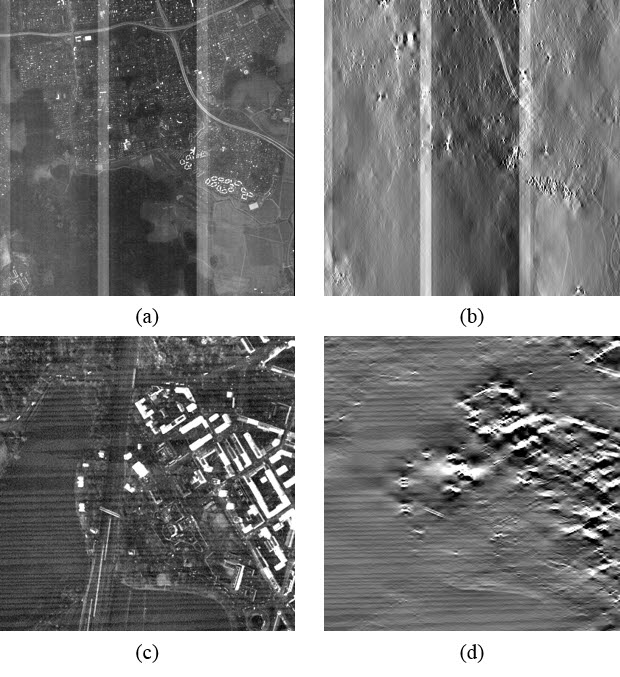}
\vspace{-0.2cm}
\caption{Examples of satellite images and wavelet subband images. (a) Image with vertical stripe noise, and (b) vertical wavelet subband image from (a). (c) Image with wave noise, and (d) horizontal wavelet subband image from (c).}
\label{figure2}
\end{figure}

For example, for the case of  vertical stripe noise in Fig.~\ref{figure2}(a), the wavelet subband images are obtained by wavelet recomposition
using  the vertical detail subbands $\{HL_{i}\}_{i=1}^{9}$ and zeroing out the other bands.  This generates an wavelet subband image  shown in Fig.~\ref{figure2}(b), which clearly
 shows the noise signals without too much of underlying structures of the scene. 
 In the case of images corrupted with the
 wave noise as shown in Fig.~\ref{figure2}(c),
 % consisting of horizontally structured patterns, 
 we can find that the subbands $\{LH_{i}\}_{i=1}^{6}$ contains the most of the noises, so we use these band
 to obtain the wavelet subband images in Fig.~\ref{figure2}(d). 

 After a denoising network remove noise patterns in noisy wavelet subband images, clean output images can be acquired by subtracting predicted noise patterns from noisy images.

\subsection{Wavelet Subband Cycle-consistent Adversarial Network}

As for an unsupervised denoising network for the wavelet subband images,
we use the cycleGAN architecture. More details are provided in the following.
%We propose a wavelet subband cycle-consistent adversarial network (WavCycleGAN) for the denoising satellite images, in which the network is based on a cycle-consistent adversarial network \cite{Reference:zhu2017unpaired} and uses wavelet subband images. Our goal is denoising vertical stripe noise patterns and wave noise patterns. 

\begin{figure}[!t]
\centering
\includegraphics[width=1\linewidth]{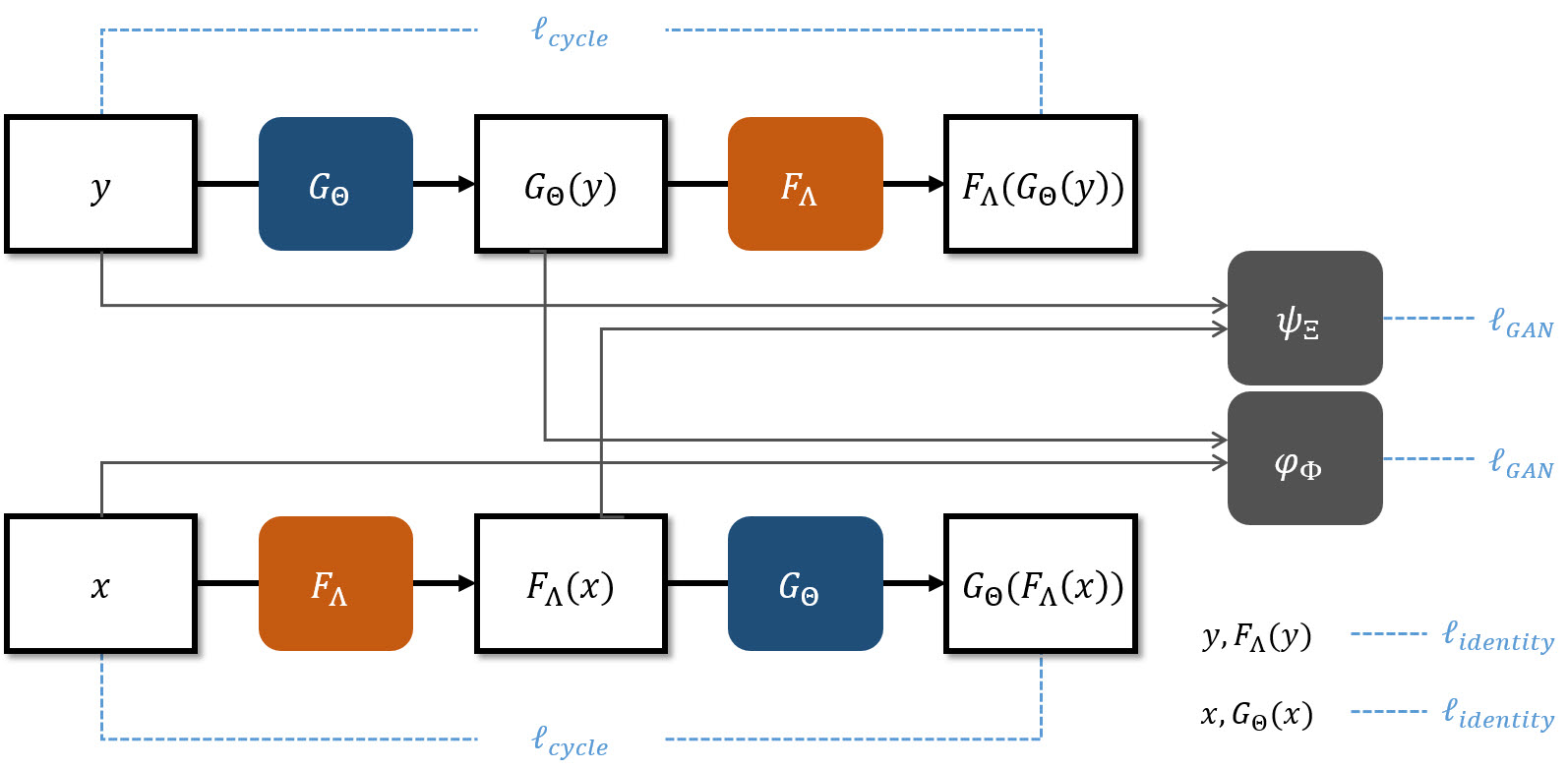}
\vspace{-0.5cm}
\caption{Architecture of our WavCycleGAN for denoising satellite images. $x$ and $y$ are wavelet subband images from the clean  domain $\Xc$ and the noisy  domain $\Yc$, respectively. The full objective consists of the adversarial loss $\ell_{GAN}$, cycle-consistency loss $\ell_{cycle}$, and identity loss $\ell_{identity}$. By minimizing the full objective function with respect to generators ($G_\Theta$ and $F_\Lambda$) and discriminators ($\psi_\Xi$ and $\varphi_\Phi$), the denoising network $G_\Theta$ can be trained in an unsupervised manner using wavelet subband images.}
\label{figure3}
\end{figure}

\subsubsection{Loss Formulation}

We consider two domains: clean domain ($\Xc$) and noisy domain ($\Yc$). The clean domain contains wavelet subband images without noise patterns, while the noisy domain consists of wavelet subband images with noise patterns.  The two data domains are composed
of data that are not matched to each other.
$P_\Xc$ and $P_\Yc$ are probability distributions of the clean  domain and noisy domain, respectively. $y$ is a sample from the noisy wavelet subband image distribution, and $x$ is a sample from the clean wavelet subband image distribution.
As shown in Fig.~\ref{figure3},
$G_\Theta:\Yc\mapsto \Xc$ is the generator parameterized with
$\Theta$, which convert a noisy wavelet subband image to a clean wavelet subband image;  the generator $F_\Lambda:\Xc\mapsto \Yc$ is a generator parameterized by $\Lambda$ which generates a synthetic noisy wavelet subband image from a clean wavelet subband image. $\psi_\Xi$  is a adversarial
discriminator parameterized by $\Xi$ that distinguishes synthetic noisy wavelet subbands from real noisy wavelet subbands.
Similarly, $\varphi_\Phi$ is adversarial discriminators that distinguish denoised wavelet subband images  from real clean wavelet subband images. 

To train the wavelet subband cycle-consistent adversarial network for the denoising problem, our objective consists of three loss functions: adversarial loss $\ell_{GAN}$, cycle-consistency loss $\ell_{cycle}$, and identity loss $\ell_{identity}$.
%The adversarial loss is to train generators and discriminators in the denoising network. 
Specifically, the typical adversarial loss for the generator $G_\Theta$ and the discriminator $\varphi_\Phi$ is as follows:

\small
\begin{align}
\begin{split}
\ell_{GAN}(\Theta,\Phi) & =\mathbb{E}_{x \sim P_\Xc}[\log \varphi_\Phi(x)] \\
& \quad + \mathbb{E}_{y \sim P_\Yc}[\log (1-\varphi_\Phi(G_\Theta(y)))],
\label{eqn:gan_loss}
\end{split}
\end{align}
\normalsize

\noindent To train $G_\Theta$ and $\varphi_\Phi$, we need to solve the min-max problem as follows:

\small
\begin{align}
\min_{\Theta}\max_{\Phi}\ell_{GAN}(\Theta,\Phi)
\label{eqn:min-max_gan_loss}
\end{align}
\normalsize

\noindent The least squares GAN (LSGAN) \cite{Reference:mao2017least} uses the least square loss function instead of the cross entropy loss to overcome the problem of vanishing gradients. We adopted LSGAN for the min-max problem as follows:

\small
\begin{align}
\label{eqn:lsgan_loss}
&\min_{\Theta}\mathbb{E}_{y \sim P_\Yc}[(\varphi_\Phi(G_\Theta(y))-1)^2], \\
&\min_{\Phi}\frac{1}{2}\mathbb{E}_{x \sim P_\Xc}[(\varphi_\Phi(x)-1)^2] + \frac{1}{2}\mathbb{E}_{y \sim P_\Yc}[\varphi_\Phi(G_\Theta(y))^2],
\end{align}
\normalsize

\noindent By solving the min-max game, $G_\Theta$ is trained to generate synthesized clean wavelet subband images from real noisy wavelet subband images and deceive the discriminator $\varphi_\Phi$, while $\varphi_\Phi$ learns to discriminate between synthesized clean wavelet subband images $G_\Theta(y)$ and real clean wavelet subband images $x$. When the networks converge, $G_\Theta$ produces realistic clean wavelet subband images, and $\varphi_\Phi$ cannot distinguish between real clean wavelet subband images and synthesized clean wavelet subband images from $G_\Theta$. The role of the  adversarial loss for $F_\Lambda$ and $\psi_\Xi$ is similar to that of $G_\Theta$ and $\varphi_\Phi$.

The generators $G_\Theta$ and $F_\Lambda$ can be trained to generate realistic clean wavelet subband images by minimizing the adversarial loss. However, using only the adversarial loss may cause artificial features due to the mode collapsing problem. We used cycle-consistency loss to impose one-to-one mapping between input images and output images to reduce artifacts and to maintain important features other than noise components. The cycle-consistency loss is defined using the L1 norm as follows:

\small
\begin{align}
\begin{split}
\ell_{cycle}(\Theta,\Lambda) & =\mathbb{E}_{y \sim P_\Yc}[||F_\Lambda(G_\Theta(y))-y||_1] \\
& \quad + \mathbb{E}_{x \sim P_\Xc}[||G_\Theta(F_\Lambda(x))-x||_1],
\label{eqn:cycle_loss}
\end{split}
\end{align}
\normalsize

\noindent By enforcing the cycle-consistency for the networks, the generators $G_\Theta$ and $F_\Lambda$ can be inverse mappings of each other, in which important features of images can be maintained during the domain translation. 

Once the network is trained, at the inference phase, the denoiser $G_\Theta$ is only used.
However,  in many practical situations, many input images or image patches for the denoiser $G_\Theta$ may not be corrupted by the noise patterns. % is no guarantee that the input will have noise patterns
A desired generator $G_\Theta$ therefore should remove the noise pattern in the noisy wavelet subband image while maintaining the input wavelet subband images if noises are not present. Also, the desired generator $F_\Lambda$ adds the noise pattern when the input is clean, while maintaining the input image when the input has the noise pattern. This condition is often called  identity property, i.e. $G_\Theta(x) \simeq x$ and $F_\Lambda(y) \simeq y $ \cite{Reference:kang2019cycle}. To enforce this, we define the identity loss  as follows:

\small
\begin{align}
\begin{split}
\ell_{identity}(\Theta,\Lambda) & =\mathbb{E}_{x \sim P_\Xc}[||G_\Theta(x)-x||_1] \\
& \quad + \mathbb{E}_{y \sim P_\Yc}[||F_\Lambda(y)-y||_1] \quad .
\label{eqn:identity_loss}
\end{split}
\end{align}
\normalsize
The overall loss function is defined using $\ell_{GAN}$, $\ell_{cycle}$, and $\ell_{identity}$ as follows:

\small
\begin{align}
\begin{split}
\ell(G_\Theta,F_\Lambda,\psi_\Xi,\varphi_\Phi) & = \ell_{GAN}(\Theta,\Phi)  + \ell_{GAN}(\Lambda,\Xi) \\
& \quad + \lambda \ell_{cycle}(\Theta,\Lambda)  + \gamma \ell_{identity}(\Theta,\Lambda),
\label{eqn:overall_loss}
\end{split}
\end{align}
\normalsize

\noindent where $\lambda$ and $\gamma$ are hyperparameters for controlling the ratio of the losses between $\ell_{GAN}$, $\ell_{cycle}$, and $\ell_{identity}$. To train the WavCycleGAN for the denoising problem, we aim to optimize the following the problem:

\small
\begin{align}
\min_{\Theta,\Lambda}\max_{\Xi,\Phi}\ell(G_\Theta,F_\Lambda,\psi_\Xi,\varphi_\Phi),
\label{eqn:min-max_overall_loss}
\end{align}
\normalsize

The corresponding  architecture is given in Fig.~\ref{figure3}.
Notice that our WavCycleGAN uses wavelet subband images consisting of selected directional subbands, while standard CycleGAN use typical images. By considering prior knowledge of noise patterns, the networks easily learn the properties of structured noise patterns and show improved performance compared to results of learning typical images, as will shown in the experimental section.

\subsection{Reconstruction Flow for Specific Noise Patterns}

\subsubsection{Vertical Stripe Noise Removal}

We found that vertical stripe noise patterns are distributed globally in images. Therefore, the networks need to see the image in full resolution and capture the overall trend of the stripe patterns to learn the relationship between clean and noisy images. However, the full resolution of a test scene with vertical stripe noise patterns is 3000 $\times$ 3000 pixels, which requires huge GPU memory and high computational cost. %, so the actual training and implementation are difficult due to the limited capacity of the hardware. 
To mitigate this problem, we used a prior knowledge that vertical stripe noise patterns are similar in the vertical direction. Accordingly, we applied  downsampling along the vertical direction of the wavelet subband images by a factor of 32. By using downsampled images, the networks can be trained using images with global appearance of the stripe pattern and the computational costs can be also reduced. To train the WavCycleGAN for the removal of vertical stripe noise, we used randomly cropped patches with a size of 2048 $\times$ 32 pixels from downsampled vertical wavelet subband images. 

\begin{figure}[!hbt]
\centering
\includegraphics[width=1\linewidth]{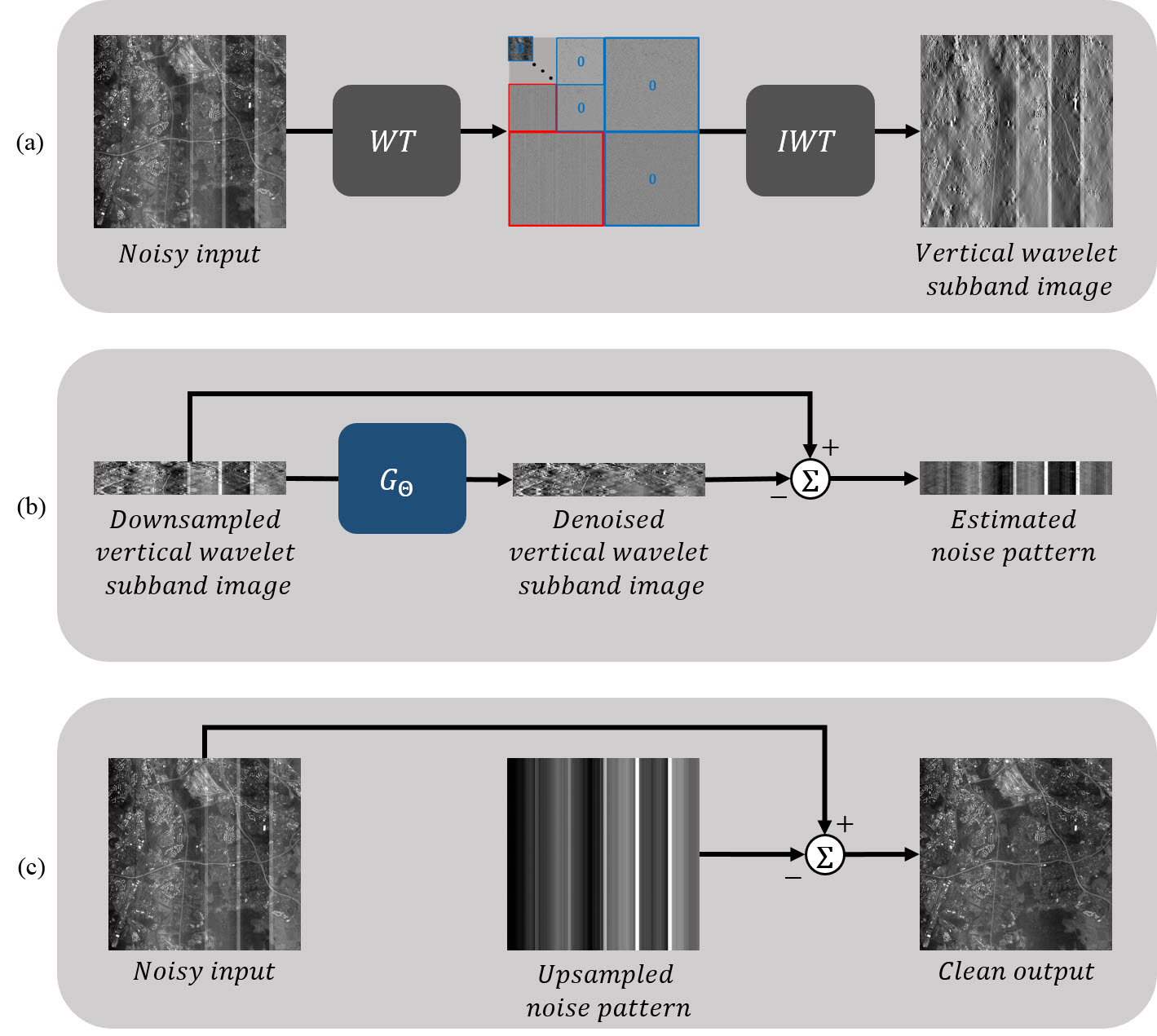}
\vspace{-0.5cm}
\caption{Overall flow of our denoising method for the vertical stripe noise. (a) A process of making a vertical wavelet subband image. WT denotes a wavelet transform, and IWT refers to an inverse wavelet transform. The red-colored subbands are only used for wavelet recomposition.
(b) A process of estimating a noise pattern. (c) The process of reconstructing a clean output from a noisy input and an upsampled noise pattern.}
\label{figure4}
\end{figure}

Fig.~\ref{figure4} shows the overall flow of our denoising method for the vertical stripe noise. First, the vertical wavelet subband image is generated by using db3 wavelet transform at the 9 decomposition level from the noisy input. When we apply the inverse wavelet transform, we preserve the coefficients of vertical bands (HL bands) and make the coefficients of the other bands (LL, LH, and LL bands) to zero. Second, the generator $G_\Theta$ removes the noise pattern of the downsampled vertical wavelet subband image. The estimated noise pattern can be acquired by subtracting the denoised wavelet subband image from the downsampled wavelet subband image. Finally, the clean output can be reconstructed by subtracting the upsampled noise pattern from the noisy input.

\begin{figure}[!t]
\centering
\includegraphics[width=1\linewidth]{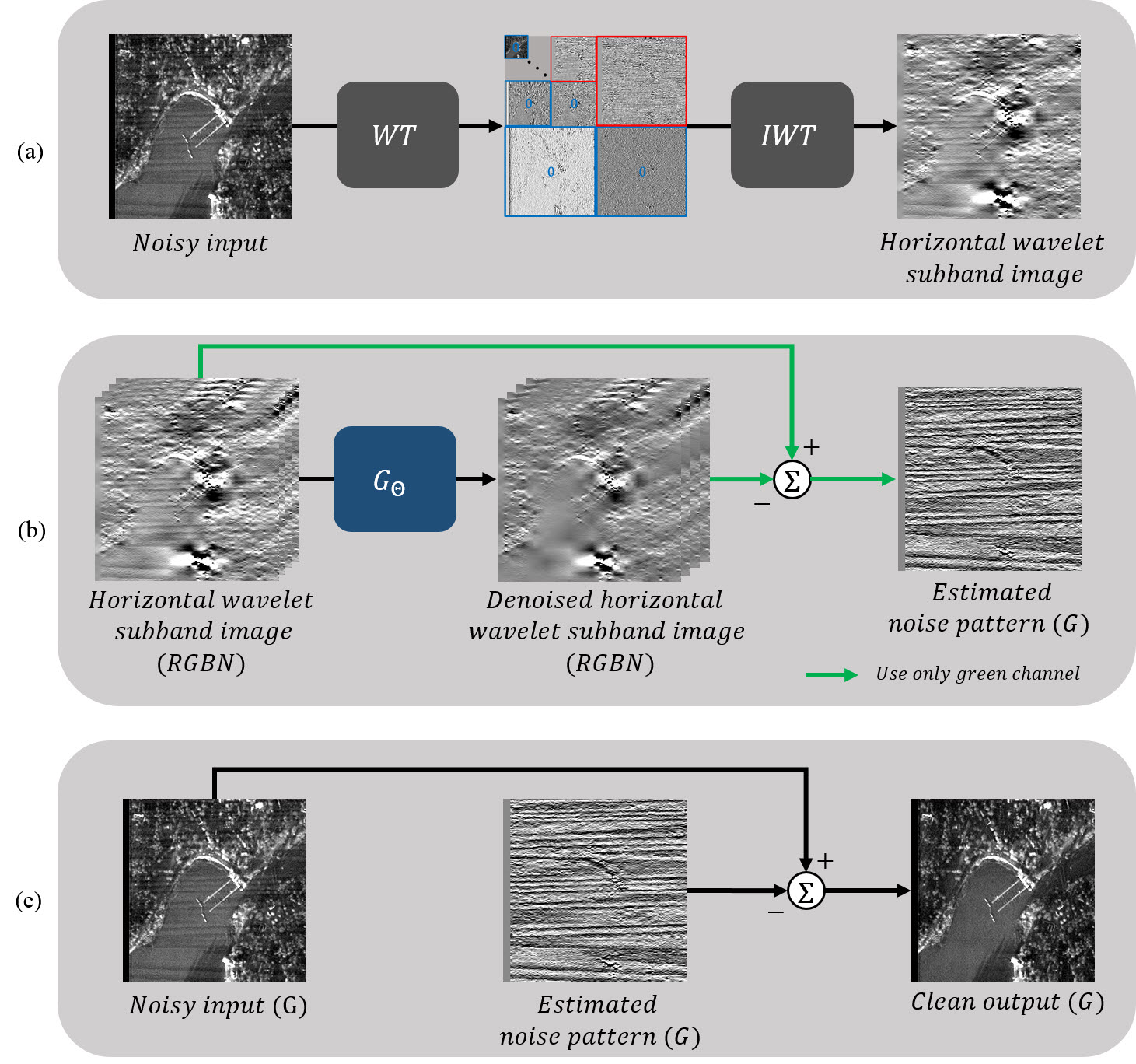}
\vspace{-0.5cm}
\caption{Overall framework of our denoising method  for the wave noise. (a) A procedure for generating a horizontal wavelet subband image. WT and IWT denote a wavelet transform and an inverse wavelet transform, respectively. We created horizontal wavelet subband image for each spectral-band image. The red-colored subbands are only used for wavelet recomposition. (b) A process of estimating a noise pattern. The predicted noise pattern can be calculated by using wavelet subband images of the green band. (c) A process of reconstructing a clean output image from a noisy image and a predicted noise pattern.}
\label{figure5}
\end{figure}

After the training process, we can only use the denoiser $G_\Theta$ in the inference stage for the denoising problem. Specifically, the noise pattern can be calculated by subtracting the wavelet subband image reconstructed by the generator $G_\Theta$ from the noisy wavelet subband image. However, due to the downsampling of the input image by the factor of 32,  the resolution of the estimated noise pattern differs from that of the input image. To increase the resolution of the noise pattern, we applied the upsampling process to the estimated noise pattern. The final reconstruction result can be obtained by subtracting the upsampled noise pattern from the noisy input.

\subsubsection{Horizontal Wave Noise Removal}
We utilized spatially registered RGBN images for the removal of wave noise. The use of spatially registered RGBN images makes the network to use the spatial correlation between multi-channel images, which improves reconstruction performance. To train the WavCycleGAN for the removal of wave noise, we used randomly cropped RGBN image patches with the size of 128 $\times$ 128 pixels from horizontal wavelet subband images. 

Fig.~\ref{figure5} shows the overall framework of our denoising method for the wave noise. The first step is generating horizontal wavelet subband images. We produced horizontal wavelet subband images for each channel image. Next, the reconstructed horizontal wavelet subband image can be acquired by using the generator $G_\Theta$. Since wave noise patterns are only present in green channel images, we calculated the noise pattern by subtracting the predicted horizontal wavelet subband image from the noisy horizontal wavelet subband image of the green channel. Finally, the clean output image can be acquired by subtracting the estimated noise pattern from the noisy input image of the green channel.

At the inference stage, the algorithms are  applied by overlapping patch images by half to avoid blocking artifacts.
% from the full scene and reconstruct them using the generator $G_\Theta$.
The reconstructed full scene are then acquired by assembling only center parts of reconstructed patches. Specifically, we reconstruct 128 $\times$ 128 pixel patches, and use only the center parts of patches with the size of 64 $\times$ 64 pixels.

\section{Methods}

\subsection{Data Set}
\subsubsection{Real Noisy Data}
In this study, we utilized multi-spectral images from a high-resolution satellite. The multi-spectral images are composed of multi-spectral images from red (R), green (G), blue (B), near-infrared (N) imaging sensors. 
The data set are corrupted by either stripe noise or wave noise depending on the type of satellite imagery.

In order to develop a denoising algorithm for the vertical stripe noises that are mainly contained in the B channel, we used blue channel images from 14 scenes with a size of 6000 $\times$ 3000 pixels. This is because the multi-spectral data were provided by the data distributor (Korea Aerospace Research Institute: KARI) without registration, so we could not use them all together.
For the removal of wave noise, we used 16 RGBN images in which each band image is of the size of 6000 $\times$ 6000 pixels. In this case,
the RGBN data were distributed by KARI with the image registration, so we aim to exploit the multi-spectral band redundancy. In our data, only green channel images have wave noise patterns. 

For every scene, the upper part was used as training data and the lower part was used as test data. 
In real situation, it is difficult to have completely clean images. 
To get clean images, we applied the conventional model-based reconstruction methods and used the resulting processed images as clean image reference for training the denoising network in unsupervised set-ups.

\subsubsection{Synthetic Noisy Data}
The development of  denoising algorithms with real samples leads to the difficulty of quantitative evaluation, as  there is no clean ground-truth corresponding to a noisy image. Without ground truths, it is not possible to calculate quantitative metrics for the image reconstruction such as the peak signal-to-noise ratio (PSNR) and structural similarity index metric (SSIM) \cite{Reference:wang2004image}. For quantitative evaluation of the algorithm, we therefore  added synthesized noise to relatively clean data. To obtain a ground-truth image for quantitative evaluation,
% synthesize noise, 
 we obtain synthetic noise patterns by subtracting the conventional model-based reconstruction results from the noisy images. Then, the synthetic noise patterns are added to the ground-truth image to generate synthetic noisy image data.

For the task of vertical stripe noise removal, we generated eight synthetic image pairs with a size of 3000 $\times$ 3000 pixels that are not used for the training of the denoising network. 
For the wave noise removal, we generated four synthetic image pairs with a size of 3000 $\times$ 6000 pixels  which are never used in the training data. When we added synthesized noise to green channel images, spatial correlation with other channels (red, blue, and near-infrared bands) were found different from real data. For instance, if the pixel values of the synthesized noisy image exceed the specified interval (e.g. [0, 65535]), values outside the interval are need to be clipped, which leads to an incorrect spatial correlation with other bands. 
%However, our method uses the spatial correlation between multi-channel images for effective noise removal. 
Therefore, for a quantitative evaluation of horizontal wave images, we only compared results of neural network using green band images.

 It is remarkable that these synthetic data are only used at the inference phase.
% , and our neural network is trained
%using only unmatched real data set.

%To evaluate the performance of our method, we used 4 image pairs. 

\subsection{Implementation Details}

\subsubsection{The Architecture of Generators and Discriminators}
For the generators $G_\Theta$ and $F_\Lambda$ in our denoising model, we used the tight-frame U-net \cite{Reference:han2018framing} structure with the skip connection between the input and output nodes. The tight-frame U-net uses wavelet decomposition and concatenation instead of conventional pooling and unpooling layers in order to satisfy the frame condition so that the networks effectively reconstruct high frequency components \cite{Reference:ye2018deep}. Furthermore, by adding the skip-connection between input and output nodes, we exploited the residual learning scheme, which is effective for denoising \cite{Reference:zhang2017beyond}. We also replaced batch normalization layers \cite{Reference:pmlr-v37-ioffe15} with instance normalization layers \cite{Reference:ulyanov2016instance} which is known for improving the quality of image generation.
The discriminators $\varphi_\Phi$ and $\psi_\Xi$ are constructed based on the structure of PatchGAN \cite{Reference:isola2017image}, which penalizes image patches to capture the texture and style of images. We used PatchGAN consisting of five convolutional layers and the fully connected layer with the instance normalization.

\begin{figure*}[!hbt]
\centering
\includegraphics[width=1\linewidth]{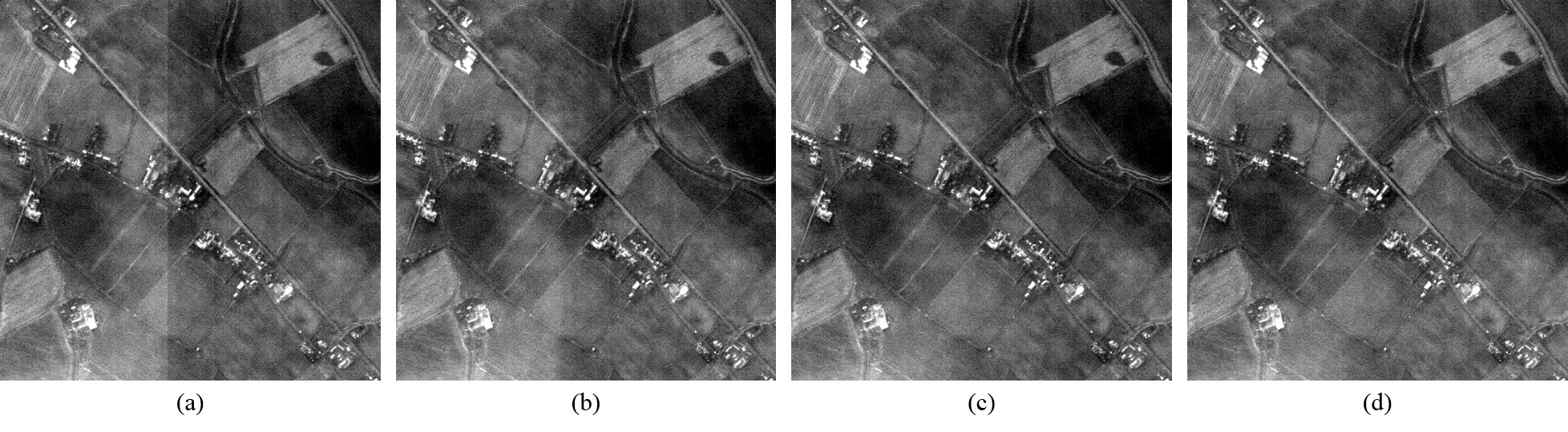}
\vspace{-0.8cm}
\caption{Results of the vertical stripe noise removal in the first scene (agricultural area): (a) noisy image, and results of (b) image-domain CycleGAN, (c)  WavCycleGAN, and (d) the conventional model-based approaches, respectively.}
\label{figure6}
\end{figure*}

\begin{figure*}[!hbt]
\centering
\includegraphics[width=1\linewidth]{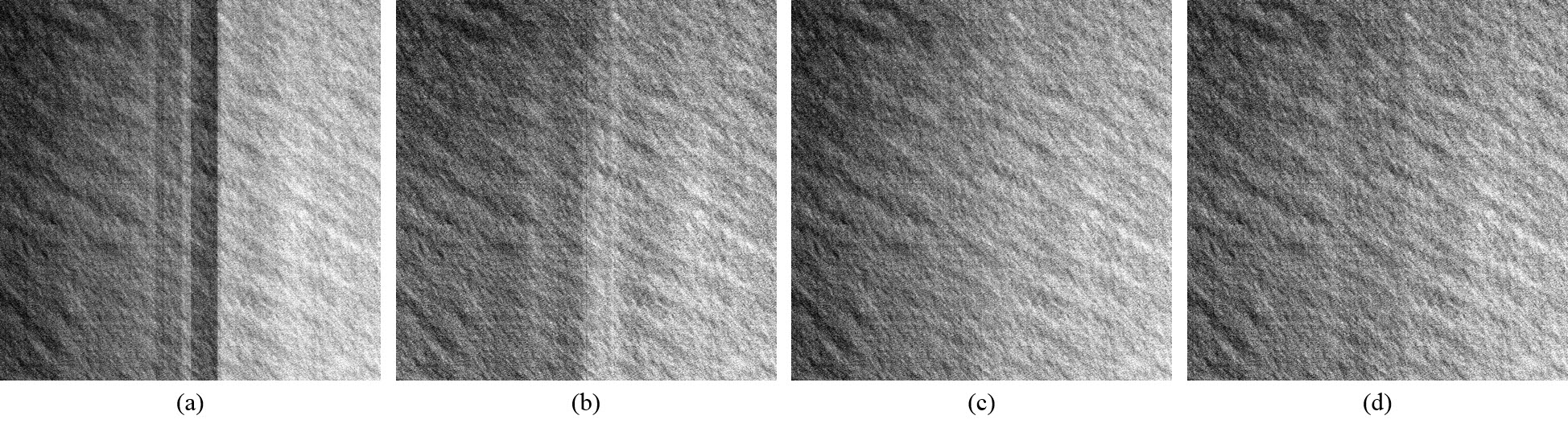}
\vspace{-0.8cm}
\caption{Results of the vertical stripe noise removal in the second scene (cloud area): (a) noisy image, and  results of (b) image-domain CycleGAN, (c)  WavCycleGAN, and (d) the conventional model-based approaches, respectively.}
\label{figure7}
\end{figure*}

\subsubsection{Training details}
For the success of supervised deep neural networks, a large amount of training data is often required. In addition, the variety of samples is an important factor. However, in many situations, the large number of data sets are not available and we consider such extreme situation to validate the advantages of our network.

Specifically, due to the security issues, our training data had fewer than 20 scenes for the development of the noise removal algorithm. To mitigate the deficiency of training data, we utilized image patches cropped from the full scenes. Specifically, we randomly cropped image patches with the size of 2048 $\times$ 32 pixels from the downsampled vertical wavelet subband images for the algorithm of denoising the vertical stripe noise. For the denoising method of the wave noise, we utilized image patches with the size of 128 $\times$ 128 pixels randomly cropped from the horizontal wavelet subband images.
We also used data augmentation strategies such as horizontal flipping and vertical flipping. The use of image patches, which are cropped randomly at each iteration in the training phase, increases the variety of the samples and a large number of training images can be acquired.

For unsupervised training, we randomly shuffle image pairs so that the network use unmatched data for the training. Our WavCycleGAN was trained by solving the optimization problem (\ref{eqn:min-max_overall_loss}) with $\lambda=10$ and $\gamma=5$. The size of mini-batch was 1. Adam optimizer \cite{Reference:kingma2014adam} was used to optimize the loss function with $\beta_1=0.5$ and $\beta_2=0.999$. The network was trained for 200 epochs. The initial learning rate was $2\times10^{-3}$ during the first 100 epochs, and gradually decreased to 0 through the last 100 epochs. The implementation of our method was based on PyTorch library \cite{Reference:paszke2019pytorch} using a NVIDIA GeForce GTX 1080 Ti GPU.

\subsection{Comparative Methods}

To evaluate the performance of  vertical stripe noise removal, we compared our method (WavCycleGAN) with various methods. 
Specifically,  in order to investigate the effectiveness of learning wavelet subband images, we also generate reconstruction results using standard image domain cycleGAN (CycleGAN) that does not utilize any directional decomposition using wavelet transform.
We also compared the conventional model-based approach. 
The conventional model-based method was based on a prior model  of 
the strip noises. The conventional model also exploited a moment matching approach \cite{Reference:gadallah2000destriping}, in which the sensors are assumed to have a linear relationship with each other.  Specifically, the model-based algorithm estimates the initial points of vertical stripes, and use edge information to calculate the positions of the noise. The initial points of stripes are calculated based on information of sensors. Using edge information of the input, homogeneous areas are selected and the start and end points of the noise are calculated based on the initial points of the noise in the homogeneous area. Vertical stripe noise patterns are estimated by subtracting the average value of the areas near the vertical pattern from the average value of vertical stripe area.

For the case of  wave noise removal, %we also compared our method with various methods. 
%Since  
multi-spectral images (RGBN bands) are registered for the case of wave noise, so we compared our results (WavCycleGAN$_{RGBN}$) with various variations to verify the benefits of our framework. Specifically, we generated comparative reconstruction
results by the image domain cycleGAN using green channel images (CycleGAN$_{G}$), wavelet subband domain cycleGAN using green channel images (WavCycleGAN$_{G}$), and the image domain cycleGAN using multi-spectral bands (CycleGAN$_{RGBN}$).
We also compared the conventional model-based approach for the wave noise. The conventional method assumed that the panchromatic image can be represented by a linear combination of multi-spectral band images with the least square regression coefficients\cite{Reference:price1987combining}. Clean green band images are then calculated using the relationship of the panchromatic image and the multi-spectral images according to the block-based scheme.

\begin{figure*}[!hbt]
\centering
\includegraphics[width=1\linewidth]{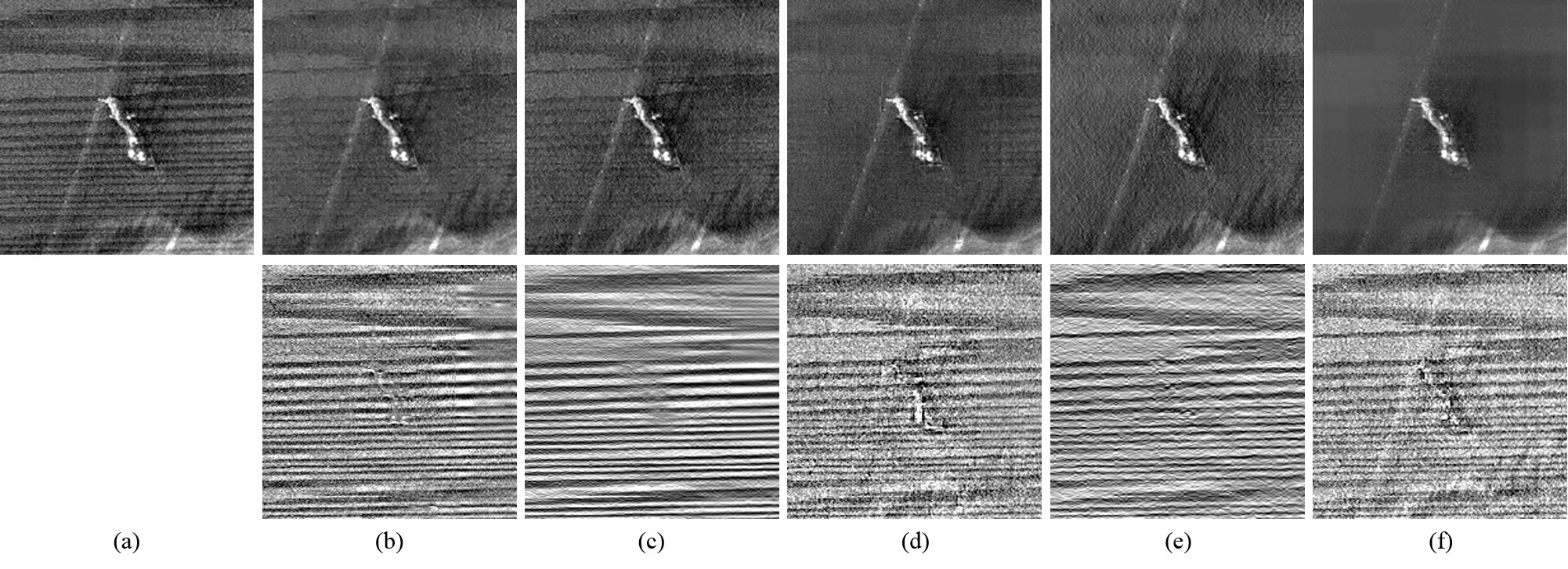}
\vspace{-0.8cm}
\caption{Results of the wave noise removal (first row) and difference images (second row) in the third scene (ocean): (a) noisy image, and results of (b) CycleGAN$_{G}$,  (c) WavCycleGAN$_{G}$, (d) CycleGAN$_{RGBN}$, (e) the proposed WavCycleGAN$_{RGBN}$, and (f) the  conventional model-based approach.}
\label{figure8}
\end{figure*}

\begin{figure*}[!hbt]
\centering
\includegraphics[width=1\linewidth]{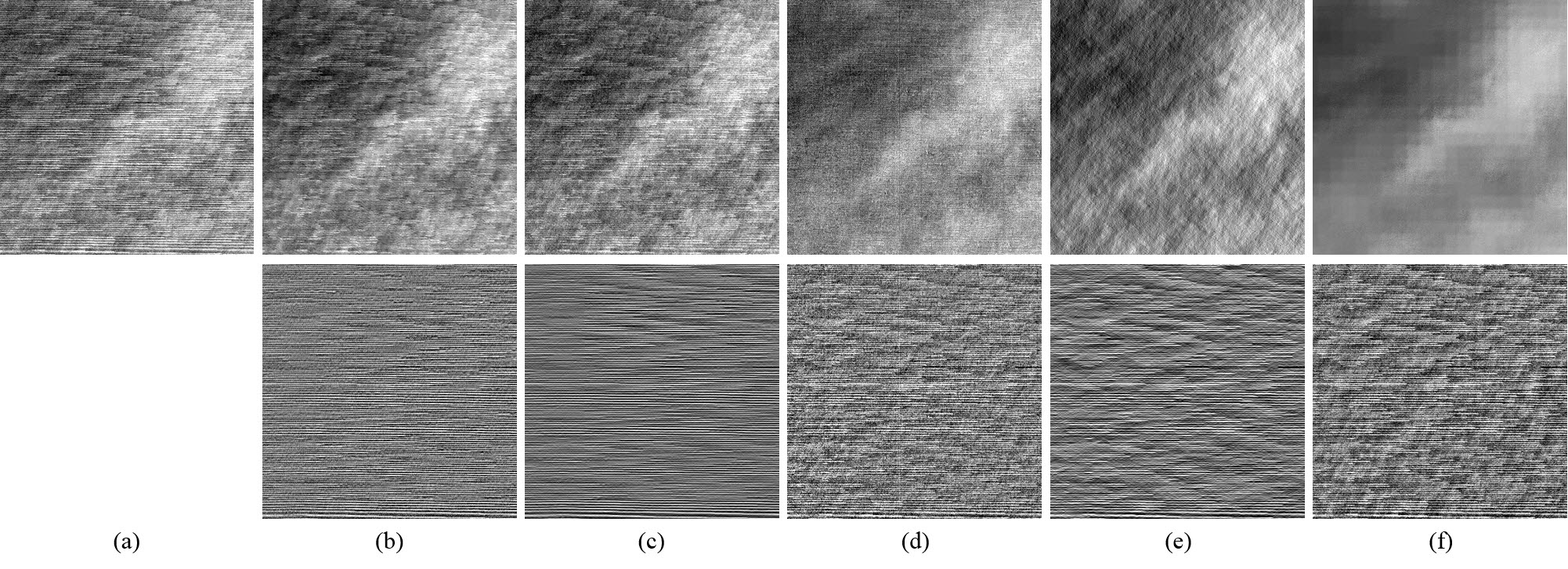}
\vspace{-0.8cm}
\caption{Results of the wave noise removal (first row) and difference images (second row) in the fourth scene (cloud): (a) noisy image, and results of (b) CycleGAN$_{G}$,  (c) WavCycleGAN$_{G}$, (d) CycleGAN$_{RGBN}$, (e) the proposed WavCycleGAN$_{RGBN}$, and (f) the conventional model-based approach.}
\label{figure9}
\end{figure*}

\section{Experimental Results}

\subsection{Real Experiments}

\subsubsection{The removal of vertical stripe noise}
To evaluate the performance of our denoising method for the vertical stripe noise, we visually inspected our results (WavCycleGAN) and compared them with other methods.

Fig.~\ref{figure6} and Fig.~\ref{figure7} show denoising results for the image patches from the first scene (agricultural area) and second scene (cloud area), respectively. The reason we chose two drastically different scenes is to validate the generalization capability of our neural network.
For Fig.~\ref{figure6}, we selected image patches with the size of 400 $\times$ 400 pixels showing significant vertical stripe noise from the first scene. The image patch of size 800 $\times$ 800 pixels was cropped from the second scene for Fig.~\ref{figure7}. 
As shown in figures, our results of learning wavelet subband images (WavCycleGAN) effectively remove vertical stripe noise, while results of the image domain cycleGAN (CycleGAN) fail to capture the noise patterns. In particular, our method successfully removed noises without  affecting high frequency components such as edges and textures. Compared with the conventional model-based results, our results show improved performance in terms of image homogeneity. For instance, in Fig.~\ref{figure7}(d), the middle part of the conventional model-based result shows image inhomogeneity, while our method shows a homogeneous denoising result.

\begin{figure*}[!hbt]
\centering
\includegraphics[width=1\linewidth]{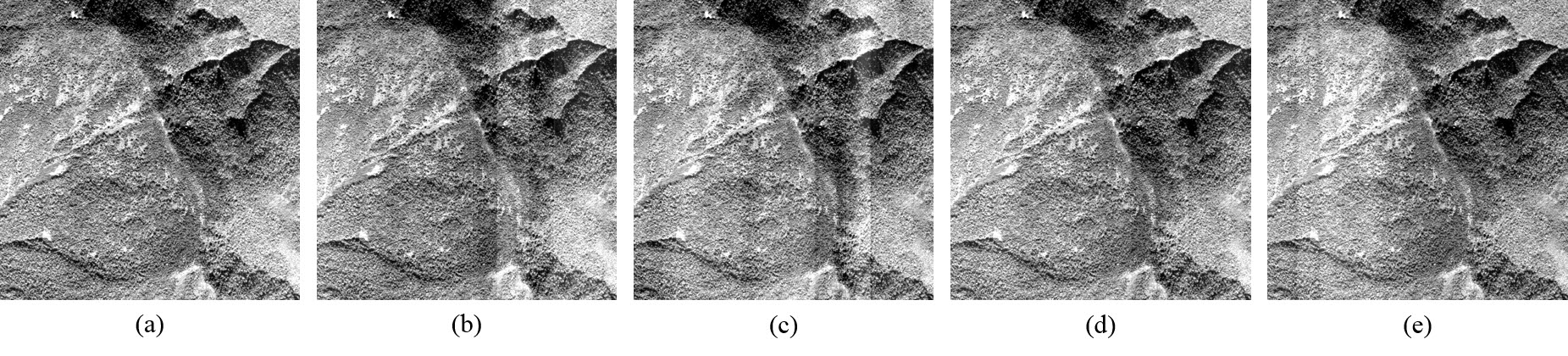}
\vspace{-0.8cm}
\caption{Results of the synthetic vertical stripe noise removal in the scene 8 (mountain area) listed in Table~\ref{table1}. (a) Ground truth image,  (b) noisy image, and results of (c)  CycleGAN, (d) our WavCycleGAN, and  (e) the conventional model-based approach.}
\label{figure10}
\end{figure*}

\begin{figure*}[!hbt]
\centering
\includegraphics[width=1\linewidth]{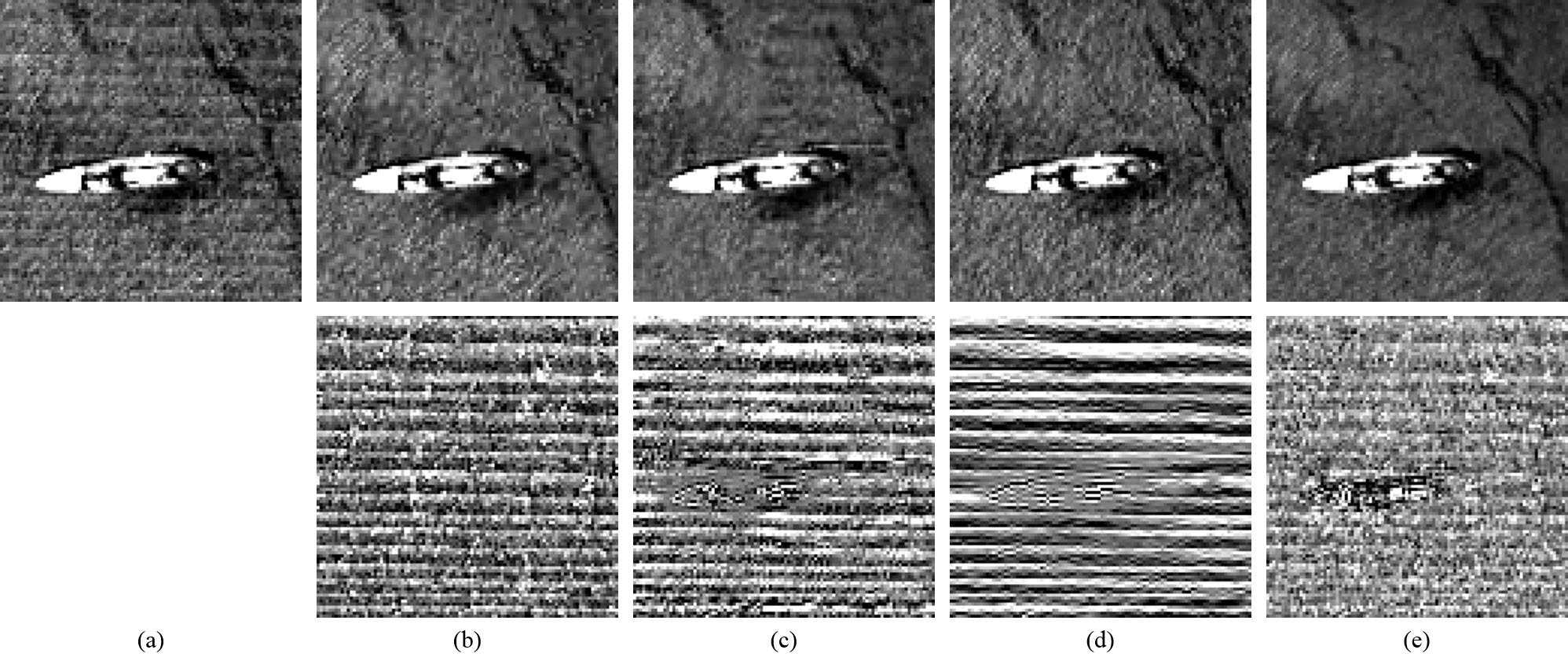}
\vspace{-0.8cm}
\caption{Results of the synthetic wave noise removal (first row) and difference images (second row) in the scene 2 (ocean area) listed in Table~\ref{table2}.
 (a) Noisy image,  (b) ground truth image,  and results of (c) CycleGAN$_{G}$, (d)  WavCycleGAN$_{G}$, and (e) the conventional model-based approach.}
\label{figure11}
\end{figure*}

\subsubsection{The removal of wave noise}

%For the qualitative evaluation of our method for the removal of the wave noise, we compared our results (WavCycleGAN$_{RGBN}$) with the results of the conventional model-based approach. Since our results are from learning multi-spectral band wavelet subband images, we also conducted ablation studies to verify the benefits of our framework. Specifically, we generated comparative reconstruction
%results by the image domain cycleGAN using green channel images (CycleGAN$_{G}$),  wavelet subband domain cycleGAN using green channel images (WavCycleGAN$_{G}$), and wavelet subband domain cycleGAN using multi-spectral bands (CycleGAN$_{RGBN}$).

 Fig.~\ref{figure8} and Fig.~\ref{figure9} show results of the wave noise removal in the third scene (ocean) and the fourth scene (cloud), respectively. Again, the reason to show  two very different scenes at the test phase is to validate the generalization power of our method.
We used the image patch with the size of 200 $\times$ 200 pixels for the third scene, and the image patch of size 400 $\times$ 400 pixels for the fourth scene. We also visualized the difference images  by subtracting denoised results from the noisy images. 

As shown in Figs.~\ref{figure8} and \ref{figure9}, results of the image domain cycleGAN  using only G channel do not successfully remove wave noise. Specifically,
  we found that  CycleGAN$_{G}$ erroneously remove structural features of objects, while results of WavCycleGAN$_{G}$ preserve these high frequency features. In Fig.~\ref{figure8}(b), the difference image contains edges of the object, while only horizontal wave patterns are present in Fig.~\ref{figure8}(c).
  In addition, our experimental results show that using multi-spectral images improves performance, since
   the network can utilize the spatial correlation between the individual spectral bands for noise reduction. However, results of CycleGAN$_{RGBN}$ tend to blur edges and details of images. The difference image of Fig.~\ref{figure8}(d) shows that the multi-spectral image domain cycleGAN (CycleGAN$_{RGBN}$) removed high frequency features which are important information for the application of satellite imagery. Compared with results of CycleGAN$_{RGBN}$, our results  using WavCycleGAN$_{RGBN}$ show effective removal of noise without sacrificing high frequency components as shown in Fig.~\ref{figure8}(e).

Furthermore, in contrast to our proposed method (WavCycleGAN$_{RGBN}$),
we found that the use of multi-spectral image domain cycleGAN (CycleGAN$_{RGBN}$) often introduce unexpected
artifacts to the green band reconstruction images. For instance, in Fig.~\ref{figure9}(d), the green band reconstruction results are corrupted by other streaks that are not present in the input image. We noticed that these artifacts are from other channel images (R, B, and N bands) during the unsupervised image domain learning.
On the other hand,
by using  wavelet subband images with horizontal bands, only horizontal components can be reconstructed, while other directional components  can be retained.  Therefore, no such artifacts are observed in the proposed method.
%Also, the complexity of the problem is reduced because the networks only learn the relationship of horizontal components between clean and noisy images, so that effective denoising can be achieved. 

It is also remarkable that  although we used conventional model-based results as a clean domain in training the networks, our unsupervised learning results show improved denoising performance than the conventional methods by learning the image distribution matching rather than pair-wise matching. 
For example, the model-based approach completely blurred out the cloud image in Fig.~\ref{figure9}(f), whereas our method provides
very high resolution image reconstruction without noises.

\begin{table}[!t]
\caption{Quantitative comparison  for the vertical stripe noise removal}
\renewcommand{\arraystretch}{1.3}
\scalebox{0.85}{%
\begin{tabular}{|c|c|c|c|c|}
\hline
\multirow{2}{*}{Scene \#} & \multicolumn{4}{c|}{PSNR [dB] / SSIM}\\\cline{2-5}
& Noisy image & CycleGAN & WavCycleGAN & Model-based\\
\hline
1 & 65.73 / 0.99982 & 61.92 / 0.99956 & \textbf{66.04} / \textbf{0.99983} & 58.42 / 0.99905 \\
\hline
2 & 66.68 / 0.99985 & 64.47 / 0.99976 & \textbf{67.14} / \textbf{0.99987} & 61.04 / 0.99946 \\
\hline
3 & 67.02 / 0.99994 & 64.60 / 0.99990 & \textbf{67.93} / \textbf{0.99995} & 55.12 / 0.99908 \\
\hline
4 & \textbf{66.49} / 0.99994 & 63.70 / 0.99988 & 66.28 / \textbf{0.99994} & 63.35 / 0.99987 \\
\hline
5 & 63.95 / 0.99992 & 63.54 / 0.99991 & 64.60 / 0.99993 & \textbf{69.85} / \textbf{0.99996} \\
\hline
6 & 65.00 / 0.99991 & 62.62 / 0.99987 & 65.11 / 0.99991 & \textbf{68.52} / \textbf{0.99993} \\
\hline
7 & 66.45 / 0.99983 & 64.13 / 0.99972 & \textbf{66.83} / \textbf{0.99985} & 64.38 / 0.99979 \\
\hline
8 & 63.63 / 0.99971 & 61.93 / 0.99958 & \textbf{65.45} / \textbf{0.99981} & 64.24 / 0.99974 \\
\hline
\textbf{Average} & 65.62 / 0.99986 & 63.36 / 0.99977 & \textbf{66.17} / \textbf{0.99988} & 63.11 / 0.99961 \\
\hline
\end{tabular}}
\label{table1}
\centering
\end{table}

\subsection{Numerical simulation}
For the quantitative evaluation, we performed inferences using synthetic noisy data and calculated quantitative metrics such as PSNR and SSIM. 
%We synthesized noise pattern by subtracting the conventional model-based results from noisy images, and added those noise patterns to relatively clean data as explained in Method.
% For the reconstruction, we employed the denoiser $G_\Theta$ trained by using the data for real noise experiments.
%  We created synthetic noise and added it to relatively clean data from images that were not used during the training of the denoising network. For quantitative measures of the performance, we used two metrics: PSNR and SSIM.  

\begin{table}[!t]
\caption{Quantitative comparison  for the wave noise removal}
\renewcommand{\arraystretch}{1.3}
\scalebox{0.85}{%
\begin{tabular}{|c|c|c|c|c|}
\hline
\multirow{2}{*}{Scene \#} & \multicolumn{4}{c|}{PSNR [dB] / SSIM}\\\cline{2-5}
& Noisy image & CycleGAN$_{G}$ & WavCycleGAN$_{G}$ & Model-based\\
\hline
1 & 51.61 / 0.99275 & 53.41 / 0.99672  & 52.92 / 0.99467  & \textbf{57.66} / \textbf{0.99836} \\
\hline
2 & 53.47 / 0.99636  & 53.41 / 0.99678 & \textbf{53.94} / \textbf{0.99686}  & 49.05 / 0.99348 \\
\hline
3 & 55.94 / 0.99743 & 59.38 / 0.99894 & \textbf{59.78} / \textbf{0.99903}  & 54.74 / 0.99768 \\
\hline
4 & 62.50 / 0.99948 & 60.01 / 0.99913  & \textbf{62.63} / \textbf{0.99952}  & 53.25 / 0.99605 \\
\hline
\textbf{Average} & 55.88 / 0.99651 & 56.55 / \textbf{0.99789}  & \textbf{57.32} / 0.99752 & 53.67 / 0.99639 \\
\hline
\end{tabular}}
\label{table2}
\centering
\end{table} 

\subsubsection{Removal of vertical stripe noise}
%We utilized 8 image pairs for the evaluation of vertical stripe noise removal. 
Table~\ref{table1} lists the PSNR and SSIM values of the scenes from 8 noisy images
and the reconstruction results by the image domain cycleGAN (CycleGAN), wavelet subband domain cycleGAN (WavCycleGAN), and the model-based method. Our results from WavCycleGAN outperform the results of CycleGAN in terms of PSNR and SSIM for the all scenes. Furthermore, we observed that the mean PSNR and SSIM values of our results are highest among other methods, which confirms that our method improves the performance by using the wavelet subband image learning scheme. Fig.~\ref{figure10} illustrates results of denoising synthetic vertical noise patterns from the image patch with the size of 600 $\times$ 600 pixels. It can be seen that WavCycleGAN outperforms CycleGAN, and our method shows homogeneous image reconstruction compared to the conventional model-based approach.

\subsubsection{Removal of wave noise}
Table~\ref{table2} shows the PSNR and SSIM scores of the scenes from 4 noisy images and the reconstruction results using image-domain cycleGAN (CycleGAN$_{G}$), wavelet subband domain cycleGAN (WavCycleGAN$_{G}$), and results of the conventional model-based method. We observed that our method (WavCycleGAN$_{G}$) outperforms the existing method (CycleGAN$_{G}$) in terms of PSNR and SSIM values with the exception of Scene 1. Also, our method yields the highest average values of PSNR. Fig.~\ref{figure11} shows reconstruction results and difference images for the removal of synthetic wave noise using the image patch with the size of 100 $\times$ 100 pixels. In contrast to the results of CycleGAN$_{G}$ and the conventional model, our method effectively removes noise pattern without removing edges and detail information.
On the other hand, the model-based approach removed the details of the scene, providing a bit blurry image.

\section{Conclusion}
In this paper, we proposed the wavelet subband cycle-consistent adversarial network (WavCycleGAN) for the multi-spectral denoising in satellite imagery. The main motivation for using WavCycleGAN is that our target noise patterns are directionally structured and only directional components of the noise pattern can be reconstructed using the wavelet subband learning scheme for efficient noise removal. Furthermore, to alleviate the problem of unpaired data in practice, we trained the denoising network in an unsupervised manner. Thanks to the use of WavCycleGAN, the denoising network could be trained efficiently in an unsupervised manner. Experimental results demonstrated that our method effectively removes noise patterns without sacrificing high frequency components.

% or


%\appendices
%% you can choose not to have a title for an appendix
%% if you want by leaving the argument blank
%\section{}

%% use section* for acknowledgment
%\section*{Acknowledgment}
%%\cite{Reference:conf_typical}.

\ifCLASSOPTIONcaptionsoff
  \newpage
\fi



%\bibliographystyle{IEEEtran}
%\bibliography{Reference}

%\bibliographystyle{IEEEtran}
\bibliographystyle{ieeetran}
\bibliography{Reference}

\begin{thebibliography}{10}
\providecommand{\url}[1]{#1}
\csname url@samestyle\endcsname
\providecommand{\newblock}{\relax}
\providecommand{\bibinfo}[2]{#2}
\providecommand{\BIBentrySTDinterwordspacing}{\spaceskip=0pt\relax}
\providecommand{\BIBentryALTinterwordstretchfactor}{4}
\providecommand{\BIBentryALTinterwordspacing}{\spaceskip=\fontdimen2\font plus
\BIBentryALTinterwordstretchfactor\fontdimen3\font minus
  \fontdimen4\font\relax}
\providecommand{\BIBforeignlanguage}[2]{{%
\expandafter\ifx\csname l@#1\endcsname\relax
\typeout{** WARNING: IEEEtran.bst: No hyphenation pattern has been}%
\typeout{** loaded for the language `#1'. Using the pattern for}%
\typeout{** the default language instead.}%
\else
\language=\csname l@#1\endcsname
\fi
#2}}
\providecommand{\BIBdecl}{\relax}
\BIBdecl

\bibitem{Reference:mulla2013twenty}
D.~J. Mulla, ``Twenty five years of remote sensing in precision agriculture:
  Key advances and remaining knowledge gaps,'' \emph{Biosystems engineering},
  vol. 114, no.~4, pp. 358--371, 2013.

\bibitem{Reference:larsen2009traffic}
S.~{\O}. Larsen, H.~Koren, and R.~Solberg, ``Traffic monitoring using very high
  resolution satellite imagery,'' \emph{Photogrammetric Engineering \& Remote
  Sensing}, vol.~75, no.~7, pp. 859--869, 2009.

\bibitem{Reference:pham2011case}
H.~M. Pham, Y.~Yamaguchi, and T.~Q. Bui, ``A case study on the relation between
  city planning and urban growth using remote sensing and spatial metrics,''
  \emph{Landscape and Urban Planning}, vol. 100, no.~3, pp. 223--230, 2011.

\bibitem{Reference:voigt2007satellite}
S.~Voigt, T.~Kemper, T.~Riedlinger, R.~Kiefl, K.~Scholte, and H.~Mehl,
  ``Satellite image analysis for disaster and crisis-management support,''
  \emph{IEEE transactions on geoscience and remote sensing}, vol.~45, no.~6,
  pp. 1520--1528, 2007.

\bibitem{Reference:yuan2012hyperspectral}
Q.~Yuan, L.~Zhang, and H.~Shen, ``Hyperspectral image denoising employing a
  spectral--spatial adaptive total variation model,'' \emph{IEEE Transactions
  on Geoscience and Remote Sensing}, vol.~50, no.~10, pp. 3660--3677, 2012.

\bibitem{Reference:chang2015anisotropic}
Y.~Chang, L.~Yan, H.~Fang, and C.~Luo, ``Anisotropic spectral-spatial total
  variation model for multispectral remote sensing image destriping,''
  \emph{IEEE Transactions on Image Processing}, vol.~24, no.~6, pp. 1852--1866,
  2015.

\bibitem{Reference:zhang2013hyperspectral}
H.~Zhang, W.~He, L.~Zhang, H.~Shen, and Q.~Yuan, ``Hyperspectral image
  restoration using low-rank matrix recovery,'' \emph{IEEE Transactions on
  Geoscience and Remote Sensing}, vol.~52, no.~8, pp. 4729--4743, 2013.

\bibitem{Reference:he2015hyperspectral}
W.~He, H.~Zhang, L.~Zhang, and H.~Shen, ``Hyperspectral image denoising via
  noise-adjusted iterative low-rank matrix approximation,'' \emph{IEEE Journal
  of Selected Topics in Applied Earth Observations and Remote Sensing}, vol.~8,
  no.~6, pp. 3050--3061, 2015.

\bibitem{Reference:he2015total}
------, ``Total-variation-regularized low-rank matrix factorization for
  hyperspectral image restoration,'' \emph{IEEE transactions on geoscience and
  remote sensing}, vol.~54, no.~1, pp. 178--188, 2015.

\bibitem{Reference:du2016pltd}
B.~Du, M.~Zhang, L.~Zhang, R.~Hu, and D.~Tao, ``Pltd: Patch-based low-rank
  tensor decomposition for hyperspectral images,'' \emph{IEEE Transactions on
  Multimedia}, vol.~19, no.~1, pp. 67--79, 2016.

\bibitem{Reference:wu2017structure}
Z.~Wu, Q.~Wang, J.~Jin, and Y.~Shen, ``Structure tensor total
  variation-regularized weighted nuclear norm minimization for hyperspectral
  image mixed denoising,'' \emph{Signal Processing}, vol. 131, pp. 202--219,
  2017.

\bibitem{Reference:wang2017hyperspectral}
Y.~Wang, J.~Peng, Q.~Zhao, Y.~Leung, X.-L. Zhao, and D.~Meng, ``Hyperspectral
  image restoration via total variation regularized low-rank tensor
  decomposition,'' \emph{IEEE Journal of Selected Topics in Applied Earth
  Observations and Remote Sensing}, vol.~11, no.~4, pp. 1227--1243, 2017.

\bibitem{Reference:fan2018spatial}
H.~Fan, C.~Li, Y.~Guo, G.~Kuang, and J.~Ma, ``Spatial--spectral total variation
  regularized low-rank tensor decomposition for hyperspectral image
  denoising,'' \emph{IEEE Transactions on Geoscience and Remote Sensing},
  vol.~56, no.~10, pp. 6196--6213, 2018.

\bibitem{burger2012image}
H.~C. Burger, C.~J. Schuler, and S.~Harmeling, ``Image denoising: Can plain
  neural networks compete with {BM3D}?'' in \emph{2012 IEEE Conference on
  Computer Vision and Pattern Recognition (CVPR)}.\hskip 1em plus 0.5em minus
  0.4em\relax IEEE, 2012, pp. 2392--2399.

\bibitem{mao2016image}
X.~Mao, C.~Shen, and Y.-B. Yang, ``Image restoration using very deep
  convolutional encoder-decoder networks with symmetric skip connections,'' in
  \emph{Advances in Neural Information Processing Systems}, 2016, pp.
  2802--2810.

\bibitem{Reference:zhang2017beyond}
K.~Zhang, W.~Zuo, Y.~Chen, D.~Meng, and L.~Zhang, ``Beyond a gaussian denoiser:
  Residual learning of deep cnn for image denoising,'' \emph{IEEE Transactions
  on Image Processing}, vol.~26, no.~7, pp. 3142--3155, 2017.

\bibitem{Reference:zhang2018ffdnet}
K.~Zhang, W.~Zuo, and L.~Zhang, ``{FFDNet: Toward a fast and flexible solution
  for CNN-based image denoising},'' \emph{IEEE Transactions on Image
  Processing}, vol.~27, no.~9, pp. 4608--4622, 2018.

\bibitem{Reference:yuan2018hyperspectral}
Q.~Yuan, Q.~Zhang, J.~Li, H.~Shen, and L.~Zhang, ``Hyperspectral image
  denoising employing a spatial--spectral deep residual convolutional neural
  network,'' \emph{IEEE Transactions on Geoscience and Remote Sensing},
  vol.~57, no.~2, pp. 1205--1218, 2018.

\bibitem{Reference:chang2018hsi}
Y.~Chang, L.~Yan, H.~Fang, S.~Zhong, and W.~Liao, ``Hsi-denet: Hyperspectral
  image restoration via convolutional neural network,'' \emph{IEEE Transactions
  on Geoscience and Remote Sensing}, vol.~57, no.~2, pp. 667--682, 2018.

\bibitem{Reference:zhang2019hybrid}
Q.~Zhang, Q.~Yuan, J.~Li, X.~Liu, H.~Shen, and L.~Zhang, ``Hybrid noise removal
  in hyperspectral imagery with a spatial-spectral gradient network,''
  \emph{IEEE Transactions on Geoscience and Remote Sensing}, 2019.

\bibitem{Reference:guan2019wavelet}
J.~Guan, R.~Lai, and A.~Xiong, ``Wavelet deep neural network for stripe noise
  removal,'' \emph{IEEE Access}, vol.~7, pp. 44\,544--44\,554, 2019.

\bibitem{Reference:liu20193}
W.~Liu and J.~Lee, ``A 3-d atrous convolution neural network for hyperspectral
  image denoising,'' \emph{IEEE Transactions on Geoscience and Remote Sensing},
  vol.~57, no.~8, pp. 5701--5715, 2019.

\bibitem{Reference:shan2019hyperspectral}
W.~Shan, P.~Liu, L.~Mu, C.~Cao, and G.~He, ``Hyperspectral image denoising with
  dual deep cnn,'' \emph{IEEE Access}, vol.~7, pp. 171\,297--171\,312, 2019.

\bibitem{Reference:chang2019toward}
Y.~Chang, M.~Chen, L.~Yan, X.-L. Zhao, Y.~Li, and S.~Zhong, ``Toward universal
  stripe removal via wavelet-based deep convolutional neural network,''
  \emph{IEEE Transactions on Geoscience and Remote Sensing}, 2019.

\bibitem{Reference:goodfellow2014generative}
I.~Goodfellow, J.~Pouget-Abadie, M.~Mirza, B.~Xu, D.~Warde-Farley, S.~Ozair,
  A.~Courville, and Y.~Bengio, ``Generative adversarial nets,'' in
  \emph{Advances in neural information processing systems}, 2014, pp.
  2672--2680.

\bibitem{Reference:zhu2017unpaired}
J.-Y. Zhu, T.~Park, P.~Isola, and A.~A. Efros, ``Unpaired image-to-image
  translation using cycle-consistent adversarial networks,'' in
  \emph{Proceedings of the IEEE international conference on computer vision},
  2017, pp. 2223--2232.

\bibitem{Reference:kang2019cycle}
E.~Kang, H.~J. Koo, D.~H. Yang, J.~B. Seo, and J.~C. Ye, ``Cycle-consistent
  adversarial denoising network for multiphase coronary ct angiography,''
  \emph{Medical physics}, vol.~46, no.~2, pp. 550--562, 2019.

\bibitem{Reference:kang2017deep}
E.~Kang, J.~Min, and J.~C. Ye, ``A deep convolutional neural network using
  directional wavelets for low-dose x-ray ct reconstruction,'' \emph{Medical
  physics}, vol.~44, no.~10, pp. e360--e375, 2017.

\bibitem{Reference:rudin1992nonlinear}
L.~I. Rudin, S.~Osher, and E.~Fatemi, ``Nonlinear total variation based noise
  removal algorithms,'' \emph{Physica D: nonlinear phenomena}, vol.~60, no.
  1-4, pp. 259--268, 1992.

\bibitem{Reference:mao2017least}
X.~Mao, Q.~Li, H.~Xie, R.~Y. Lau, Z.~Wang, and S.~Paul~Smolley, ``Least squares
  generative adversarial networks,'' in \emph{Proceedings of the IEEE
  International Conference on Computer Vision}, 2017, pp. 2794--2802.

\bibitem{Reference:wang2004image}
Z.~Wang, A.~C. Bovik, H.~R. Sheikh, E.~P. Simoncelli \emph{et~al.}, ``Image
  quality assessment: from error visibility to structural similarity,''
  \emph{IEEE transactions on image processing}, vol.~13, no.~4, pp. 600--612,
  2004.

\bibitem{Reference:han2018framing}
Y.~Han and J.~C. Ye, ``Framing u-net via deep convolutional framelets:
  Application to sparse-view ct,'' \emph{IEEE transactions on medical imaging},
  vol.~37, no.~6, pp. 1418--1429, 2018.

\bibitem{Reference:ye2018deep}
J.~C. Ye, Y.~Han, and E.~Cha, ``Deep convolutional framelets: A general deep
  learning framework for inverse problems,'' \emph{SIAM Journal on Imaging
  Sciences}, vol.~11, no.~2, pp. 991--1048, 2018.

\bibitem{Reference:pmlr-v37-ioffe15}
S.~Ioffe and C.~Szegedy, ``Batch normalization: Accelerating deep network
  training by reducing internal covariate shift,'' in \emph{Proceedings of the
  32nd International Conference on Machine Learning}, 2015, pp. 448--456.

\bibitem{Reference:ulyanov2016instance}
D.~Ulyanov, A.~Vedaldi, and V.~Lempitsky, ``Instance normalization: The missing
  ingredient for fast stylization,'' \emph{arXiv preprint arXiv:1607.08022},
  2016.

\bibitem{Reference:isola2017image}
P.~Isola, J.-Y. Zhu, T.~Zhou, and A.~A. Efros, ``Image-to-image translation
  with conditional adversarial networks,'' in \emph{Proceedings of the IEEE
  conference on computer vision and pattern recognition}, 2017, pp. 1125--1134.

\bibitem{Reference:kingma2014adam}
D.~P. Kingma and J.~Ba, ``Adam: A method for stochastic optimization,''
  \emph{arXiv preprint arXiv:1412.6980}, 2014.

\bibitem{Reference:paszke2019pytorch}
A.~Paszke, S.~Gross, F.~Massa, A.~Lerer, J.~Bradbury, G.~Chanan, T.~Killeen,
  Z.~Lin, N.~Gimelshein, L.~Antiga \emph{et~al.}, ``Pytorch: An imperative
  style, high-performance deep learning library,'' in \emph{Advances in Neural
  Information Processing Systems}, 2019, pp. 8024--8035.

\bibitem{Reference:gadallah2000destriping}
F.~Gadallah, F.~Csillag, and E.~Smith, ``Destriping multisensor imagery with
  moment matching,'' \emph{International journal of remote sensing}, vol.~21,
  no.~12, pp. 2505--2511, 2000.

\bibitem{Reference:price1987combining}
J.~C. Price, ``Combining panchromatic and multispectral imagery from dual
  resolution satellite instruments,'' \emph{Remote sensing of environment},
  vol.~21, no.~2, pp. 119--128, 1987.

\end{thebibliography}
\end{document}